\documentclass[runningheads]{llncs}

\usepackage[T1]{fontenc}

\usepackage{graphicx}

\usepackage{hyperref}
\usepackage{amsmath}
\usepackage{mathtools}

\usepackage[capitalise]{cleveref}
\usepackage[inline]{enumitem}
\usepackage{multirow}
\usepackage{acronym}
\usepackage{stfloats}
\usepackage[skip=0pt]{caption}
\usepackage{etoolbox}
\usepackage{dsfont}
\usepackage{listings}
\usepackage{xcolor}
\usepackage{amssymb}
\usepackage[numbers,sort&compress]{natbib}
\usepackage{tabularx}
\usepackage{booktabs}

\usepackage{orcidlink}
\usepackage{tablefootnote}

\usepackage{hyperref}
\hypersetup{
  colorlinks   = true,    
  urlcolor     = blue,    
  linkcolor    = blue,    
  citecolor    = blue      
}

\crefname{figure}{Figure}{Figures}
\crefname{table}{Table}{Tables}
\crefname{equation}{Equation}{Equations}
\crefname{section}{Section}{Sections}
\crefname{appendix}{Appendix}{Appendices}

\crefname{enumi}{}{}

\acrodef{ToT}{Tip-of-the-tongue}
\acrodef{LLMs}{large language models}
\acrodef{LLM}{large language model}

\author{
Stan Fris\inst{1}\orcidlink{0009-0000-9300-5502} \and
Sijie Tao\inst{2}\orcidlink{0000-0002-6751-5303} \and
Makoto P. Kato\inst{3}\orcidlink{0000-0002-9351-0901} \and
Maarten de Rijke\inst{1}\orcidlink{0000-0002-1086-0202}
}

\authorrunning{Fris et al.}

\institute{
University of Amsterdam, Amsterdam, The Netherlands\\
\email{stan.fris@student.uva.nl}
\and
National Institute of Informatics, Tokyo, Japan \\
\and
University of Tsukuba, Tsukuba, Japan\\
}

\begin{document}

\title{SciToT: Multimodal Tip-of-the-Tongue Retrieval for Scientific Papers}

\maketitle              

\begin{abstract}
    \vspace{-0.5cm}
Researchers frequently encounter difficulties re-finding scientific papers when they remember aspects of the content but cannot recall reliable identifiers, such as titles or author names. This type of re-finding task, known as tip-of-the-tongue (ToT) retrieval, is particularly challenging for scientific documents, where recollections may involve textual information and visual elements, including figures, diagrams, and plots. Existing ToT retrieval benchmarks do not address scientific search. 
We introduce a multimodal benchmark for scientific ToT retrieval, comprising 108,512 open-access scientific documents across five domains and 11,499 synthetically generated queries. To construct the benchmark, we propose a scalable query elicitation pipeline that independently extracts textual and visual memory cues from source documents and uses large language models to generate queries resembling incomplete human recollections. We characterize scientific ToT information needs through an empirical analysis of publicly available re-finding requests and use the outcomes to assess the generated queries using human annotations and automated quality control. The benchmark includes lexical, dense, late-interaction, and visually grounded retrieval baselines. We  release the dataset, extracted document representations, generated memory cues, evaluation splits, and accompanying generation and retrieval pipelines under an MIT license to facilitate research on multimodal scientific paper re-finding.

\end{abstract}

\vspace{-0.8cm}
\section{Introduction} 
\ac{ToT} retrieval is a known-item search task in which users attempt to re-find an item that they have previously encountered, while not having access to exact identifiers, such as document titles or author names~\cite{arguello2021tip,bhargav2022s,he2025tip}. \ac{ToT} queries are more complex and longer than conventional keyword-based search queries, and they typically contain details about the context in which a user engaged with a work~\cite{meier2021towards}. Furthermore, such queries may contain uncertainties, relative comparisons or false memories~\cite{arguello2021tip,he2025tip}. This makes \ac{ToT} retrieval a difficult task for traditional retrieval systems~\cite{arguello2026overview,lin2023decomposing,he2025tip}. Importantly, it is a highly relevant task to resolve, as it is associated with a high level of frustration in users and can lead to decreases in productivity~\cite{elsweiler2007towards}. 

\begin{figure*}[t]
\centering
\scriptsize
\begin{minipage}{\textwidth}
    \centering
    \footnotesize\textbf{Source document:}
    \textit{Debiasing Word Embeddings with Nonlinear Geometry}~\cite{cheng2022debiasing}
\end{minipage}

\vspace{0.3em}
\hrule
\vspace{0.3em}

\begin{minipage}[t]{0.27\textwidth}
    \vspace{0pt}
    \textbf{Extracted figure}

    \vspace{0.0em}

    \centering
    \includegraphics[
        width=\linewidth,
        trim={35mm 80mm 35mm 85mm},
        clip
    ]{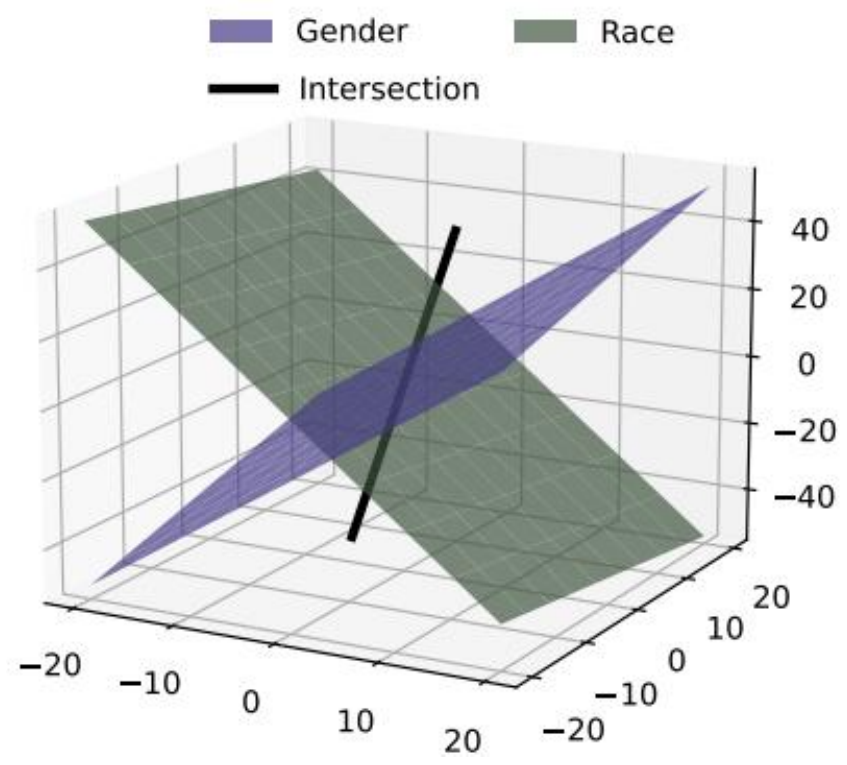}
\end{minipage}
\hfill
\begin{minipage}[t]{0.67\textwidth}
    \vspace{0pt}
    \textbf{Textual memory cues}

    \begin{itemize}[
        leftmargin=1.2em,
        itemsep=2pt,
        parsep=0pt,
        topsep=0pt
    ]
        \item Debiasing of word embeddings
        \item Discussion of intersectional biases
        \item Comparison of multiple bias mitigation methods
        \item $\cdots$
    \end{itemize}
    \vspace{-0.1em}
    \textbf{Visual memory cues}
    \begin{itemize}[
        leftmargin=1.2em,
        itemsep=2pt,
        parsep=0pt,
        topsep=0pt
    ]
        \item Three-dimensional plot
        \item Intersecting purple and green regions
        \item Black line passing through their intersection
        \item $\cdots$
    \end{itemize}
\end{minipage}
\vspace{0.3em}
\hrule
\vspace{0.2em}
\begin{minipage}{1.0\textwidth}
\textbf{Generated tip-of-the-tongue query}

\vspace{0.2em}
\itshape
I'm trying to track down a paper I came across a while
back. It involved debiasing of word embeddings and discussed
intersectional biases, comparing multiple bias mitigation
methods.
One figure that stood out was a 3D plot with intersecting
purple and green shaded regions, and a black line passing
through their intersection. Does this ring a bell?
\end{minipage}

\vspace{0.3em}
\hrule
\vspace{0.3em}
\caption{Example of the proposed tip-of-the-tongue query generation pipeline including figure, descriptive cues, and generated query. The generated query is lightly shortened for presentation; only some cues are included.}
\vspace{-0.8cm}
\label{fig:query_generation_example}
\end{figure*}

Scientific literature search is a critical task for researchers, used for collecting related work, checking previously proposed methods, and known-item search~\cite{ajith2025litsearch, li2017investigating}. Known patterns from \ac{ToT} queries persist in scientific literature search: expanded queries often contain imprecise or incorrect information, relying on attributes from remembered documents~\cite{lee2006known, lee2007incorrect}. 
Scientific literature relies on figures for supporting text, which means effective representations of visuals are essential for understanding~\cite{pramanick2024spiqa,khalighinejad2026document,zhong2025smsmo}. In \ac{ToT} retrieval, the visual component is highly relevant, as clues uttered in \ac{ToT} queries rely on remembered elements, which are often visual in nature~\cite{arguello2021tip, bhattacharyya2026unsupervised}. Our analysis of real-world \ac{ToT} queries reveals that visual elements are frequent, and can be highly distinctive elements. Hence, evaluating \ac{ToT} retrieval for scientific literature should account for remembered clues that stem from visual content. 

Despite the prevalence of \ac{ToT} style query failures in scientific literature search, no datasets exist to evaluate systems on scientific \ac{ToT} queries. Prior work in \ac{ToT} has mostly focused on casual~\cite{elsweiler2011understanding} domains such as movies, music, and games~\cite{bogers2025exploring,bhargav2022s,jorgensen2020kinda}, which can be sourced from community question-answering platforms~\cite{meier2021towards}. Importantly, for many domains, including the scientific domain, no such platforms exist with sufficient coverage for dataset construction, due to confidentiality, privacy, or data scarcity~\cite{he2025tip, meier2021towards}. 
 
In this work, we address the lack of evaluation resources for scientific \ac{ToT} retrieval by introducing a scalable approach for generating multimodal recollection queries, an example of which is shown in~\cref{fig:query_generation_example}. We extend the \ac{LLM}-based query elicitation framework proposed by \citet{he2025tip} to scientific documents and incorporate both visual and textual information. We parse PDF documents into visual and textual elements, and generate potential clues. We then generate a \ac{ToT} query based on these clues. We validate the resulting resource and establish retrieval baselines to support future research. We release the dataset, dataset construction code, retrieval code, and baseline implementations under an MIT license.

\begin{table}[t]
\centering
\caption[Comparison of \ac{ToT} and scientific retrieval benchmarks]%
{Comparison of \ac{ToT} and scientific retrieval
benchmarks. Query counts describe aggregated query set.
$^*$Paired video subset.
$^\dagger$Combined Chinese, Japanese, and Korean corpora. A dash under \#Docs indicates no fixed retrieval corpus. \#Q is short for \#Queries, Sci. for Scientific, Vis. for Visual.}
\label{tab:dataset_comparison}

\scriptsize
\setlength{\tabcolsep}{3.5pt}
\begin{tabular}{@{}l@{}rrlccc@{}}
\toprule
\textbf{Benchmark} &
\textbf{\#Docs} &
\textbf{\#Q} &
\textbf{Query Generation} &
\textbf{ToT} &
\textbf{Sci.} &
\textbf{Vis.} \\
\midrule

MS-ToT~\cite{arguello2021tip}
    & 69,132 & 1,000 & Human (collected)
    & \checkmark & -- & -- \\

TREC-ToT '25~\cite{arguello2026overview}
    & 6.4M & 622 & Mixed (collected + generated)
    & \checkmark & -- & -- \\
    
Multilingual ToT~\cite{he2026multilingual}
    & 3.42M$^\dagger$ & 20,000 & LLM (generated)
    & \checkmark & -- & -- \\
    
BLUR~\cite{sky2025browsing}
    & -- & 573 & Human (collected)
    & \checkmark & -- & \checkmark \\

ToT2MeM~\cite{bhattacharyya2026unsupervised}
    & 82.5K$^*$ & 82.5K$^*$ & Human (collected)
    & \checkmark & -- & \checkmark \\

SPIQA~\cite{pramanick2024spiqa}
    & 26,156 & 265,996 & Mixed (collected + generated)
    & -- & \checkmark & \checkmark \\

arXivDoc~\cite{khalighinejad2026document}
    & 8,210 & 547 & LLM (generated, human-verified) 
    & -- & \checkmark & \checkmark \\

\textbf{SciToT (Ours)}
    & \textbf{108,512}
    & \textbf{11,499}
    & \textbf{LLM (generated, human-verified)}
    & \checkmark
    & \checkmark
    & \checkmark \\

\bottomrule
\end{tabular}
   \vspace{-0.35cm}
\end{table}

\section{Related Work}
\vspace{-0.25cm}
We position our dataset within existing tip-of-the-tongue, multimodal, and scientific retrieval tasks. We also discuss methods for synthetic query elicitation to motivate our approach. \cref{tab:dataset_comparison} compares currently available scientific and \ac{ToT} benchmarks, illustrating that our benchmark (SciToT) addresses a gap, and generally aligns with previous work in terms of scale. 

\subsection{\acl{ToT} Retrieval}
\citet{arguello2021tip} propose \ac{ToT} retrieval as a known-item search task without reliable identifiers; they rely on conversational posts from social media. Subsequent work constructs \ac{ToT} datasets for several domains. Initial work relies strongly on domains for which datasets can be collected from online interactions, including video-games~\cite{jorgensen2020kinda}, movies and books~\cite{bhargav2022s, meier2021towards, lin2023decomposing}, and music~\cite{bhargav2023music}. Later work has increased the scale and diversity of benchmarks and methods, e.g., through a TREC task~\citep{arguello2024overview, arguello2025overview, arguello2026overview}. \citet{sky2025browsing} introduce a domain-diverse and partially multilingual \ac{ToT} corpus. \citet{he2026multilingual} introduce a large-scale multilingual corpus for \ac{ToT} Queries. \citet{bhattacharyya2026unsupervised} introduce a large-scale collection of multimodal \ac{ToT} queries including visual and textual elements from social media interactions, with a specific pairing of visual elements to textual descriptions. These works have shown that \ac{ToT} evaluation can be expanded to domains where this information need is prevalent. Our contribution is the structured generation of \ac{ToT} queries from complex multimodal documents, and the first \ac{ToT} benchmark for the domain of scientific document retrieval. 

\subsection{Scientific Literature Search}
Known-item search is especially prevalent in scientific literature search, with resources reporting up to 57\% of queries as directly looking for a known item~\cite{li2017investigating,schultheiss2020known,yeganova2018field,herskovic2007day}. Such queries are highly prone to failure, as users often cannot recall a direct identifier, resulting in many queries receiving zero results, or not retrieving the intended document~\cite{schultheiss2020known,behnert2017known}. \citet{schultheiss2020known} report that users often rely on query expansion when they cannot recall a query. Importantly, people are highly effective at remembering and recognizing images~\cite{brady2008visual}, with a strong capacity for remembering visually distinctive items~\cite{konkle2010conceptual, Isola2011memorability}. When remembering visual features of scientific papers, users are highly likely to recall semantic details of images when describing them in addition to structural elements~\cite{kramer2023features, brady2013visual,moscovitch1995distinct, borkin2015beyond}. Our benchmark builds on these key insights, which are backed by our own empirical analysis, and provides a method to generate synthetic scientific search queries that are aligned with real-world queries in style and composition. 

\subsection{Query Elicitation}
Query elicitation generates synthetic queries from ground-truth documents~\cite{rahmani2024synthetic}, supporting retrieval model training~\cite{ogilvie2003combining,phanse2024yale} and evaluation~\cite{azzopardi2006automatic,hagen2015corpus,he2025tip}. Early approaches employed statistical term-selection methods for query generation~\cite{azzopardi2007building,balog2006overview,elsweiler2011seeding,kim2009retrieval}. \Ac{LLMs} have significantly increased the possibilities for synthetic dataset generation for information retrieval, generating  labels~\cite{faggioli2023perspectives,thomas2024large}, queries~\cite{rajapakse2023improving,alaofi2023can}, or test collections~\cite{rahmani2024synthetic}. 
Recent frameworks generate queries from target documents~\cite{bonifacio2022inpars,jeronymo2023inpars}, guide generation through task-specific prompts~\cite{dai2022promptagator}, and construct synthetic test collections~\cite{rahmani2024synthetic}. \citet{he2025tip} introduce a framework for query elicitation for \ac{ToT} simulated evaluation; by fine-tuning instructions based on human-elicited queries they reach high rank correlation as well as high linguistic similarity. Later work has applied this framework for multilingual \ac{ToT}-query elicitation~\cite{he2026multilingual} and for evaluation~\cite{arguello2026overview,ntcir19tot}. Structured LLM-based query elicitation is an effective method for generating reliable and effective queries, generalizing across domains, which suggests that these methods can be effectively applied for the generation of scientific \ac{ToT} queries. What we add to this work is a structured method for query generation based on distinct elements for scientific documents, by extracting clues from figures and text and combining these in a single query. 

\section{Scientific Tip-of-the-Tongue Information Needs}
\label{sec:scitot_inf_need}


We follow the definition of a \ac{ToT} information need by~\citet{arguello2021tip} as ``an item identification task where the searcher has previously experienced or consumed the item, but cannot recall a reliable identifier.'' We consider two requirements for successful scientific ToT query elicitation:
\begin{enumerate*}[label=(\roman*)]
    \item alignment of the query with ToT information needs, assessed through an empirical characterization of re-finding requests in the scientific domain (\cref{subsec:scitot_char}).
    \item alignment of the query with the target document, which we evaluate through human annotation as well as LLM-based filtering (\cref{subsec:human_annotation,subsec:filtering})
\end{enumerate*}.

\subsection{Empirical Characterization of Scientific \ac{ToT} Queries}\label{subsec:scitot_char}
Previous work in \ac{ToT} retrieval~\cite{arguello2021tip,he2025tip,he2026multilingual} and studies of scientific literature search~\cite{schultheiss2020known,lee2006known,lee2007incorrect} identify characteristics of queries such as verbosity and uncertainty. However, to the best of our knowledge, no prior work has empirically characterized scientific \ac{ToT} queries in practice. We annotate a sample of publicly available re-finding requests collected from CQA platforms. The annotation codebook is constructed using a hybrid deductive--inductive procedure~\cite{fereday2006demonstrating}. An initial set of codes was derived from prior work on \ac{ToT} retrieval~\cite{arguello2021tip}. We then applied and refined these codes while examining the collected queries, introducing additional categories when recurring forms of recollection are not captured by the initial scheme following~\citet{bolotova2022non}. The final set of codes was determined through discussion by the authors. Named-entity associations were excluded from codes to remain consistent with prior \ac{ToT} and scientific re-finding work, which typically excludes these. Annotations were made by two annotators with scientific backgrounds, reporting annotator agreement with Cohen's $\kappa$.

\begin{table}[t]
\centering
\footnotesize
\caption{Codebook for characteristics in scientific tip-of-the-tongue queries. Freq. describes the annotated queries containing each characteristic. Cohen's $\kappa$ measures inter-annotator agreement. Asterisks $^\ast$ denote codes reused from~\citep{arguello2021tip}.}
\setlength{\tabcolsep}{3pt}
\begin{tabularx}{\textwidth}{@{}p{0.22\textwidth} X r r@{}}
\toprule
\textbf{Code} & \textbf{Definition} & \textbf{Freq.} & \textbf{$\kappa$} \\
\midrule

\multicolumn{4}{@{}l}{\textit{Textual}} \\
Research topic &
Broad scientific subject or problem studied &
93.1\% & 0.85 \\

Specific Claim &
Specific proposition, argument, or conclusion &
35.3\% & 0.74 \\

Method &
Method, model, algorithm, or analysis used &
38.2\% & 0.71 \\

Document Struct. &
General structure of information in the document &
3.9\% & 1.00 \\

Terminology &
Remembered technical term, phrase, or wording &
19.6\% & 0.75 \\

Motivation &
Practical use case or reason for the work &
4.9\% & 0.79 \\

Comparison &
Relation to another method, paper, or concept &
5.9\% & 0.65 \\

Exclusion Criteria &
Something the target is known not to be or contain &
9.8\% & 0.78 \\

Result/Finding &
General outcome or observation of the target &
28.4\% & 0.95 \\

Dataset &
Dataset, corpus, sample, or participants used &
6.9\% & 0.55 \\

\midrule
\multicolumn{4}{@{}l}{\textit{Visual}} \\
Present &
Memory that a visual element was present &
15.7\% & 1.00 \\

Graph Type &
Describes the type of graph or visual &
11.8\% & 1.00 \\

Structural &
Structure, colours, or arrangement of a visual &
7.8\% & 0.73 \\

Semantic Details &
Meaning of elements within a remembered visual &
8.8\% & 0.88 \\

\midrule
\multicolumn{4}{@{}l}{\textit{Contextual}} \\
Temporal\textbf{$^\ast$} &
When the searcher previously encountered the target &
34.3\% & 0.96 \\

Academic Activity &
Academic context in which the target was found &
4.9\% & 0.79 \\

Physical Relation &
Remembered physical or spatial relation &
41.2\% & 0.52 \\

\midrule
\multicolumn{4}{@{}l}{\textit{Social/Other}} \\
Uncertainty\textbf{$^\ast$} &
Explicit doubt or alternative recollection &
74.5\% & 0.59 \\

Previous Search\textbf{$^\ast$} &
Earlier attempts to locate or identify the target &
30.4\% & 0.68 \\

Failed Recall\textbf{$^\ast$} &
Explicit inability to remember a detail &
26.5\% & 0.65 \\

Social\textbf{$^\ast$} &
Politeness, thanks, apology, or request for help &
85.3\% & 0.77 \\

Emotion\textbf{$^\ast$} &
Emotional reaction to retrieval failure &
7.8\% & 0.73 \\

\bottomrule
\end{tabularx}
\label{tab:tot-codebook}
\vspace{-0.6cm}
\end{table}

Table~\ref{tab:tot-codebook} summarizes our annotations, together with examples and their frequencies within the collected sample of $N=51$ queries. Our coding splits docu\-ment-based (visual and structural) elements, which require the analysis of the target document, from user-based elements (Contextual and Social/ Other), which are user-specific, and may be generated without document knowledge~\cite{he2025tip}.

\section{Dataset Construction}
\begin{figure}[t]
    \centering
    \includegraphics[width=\linewidth]{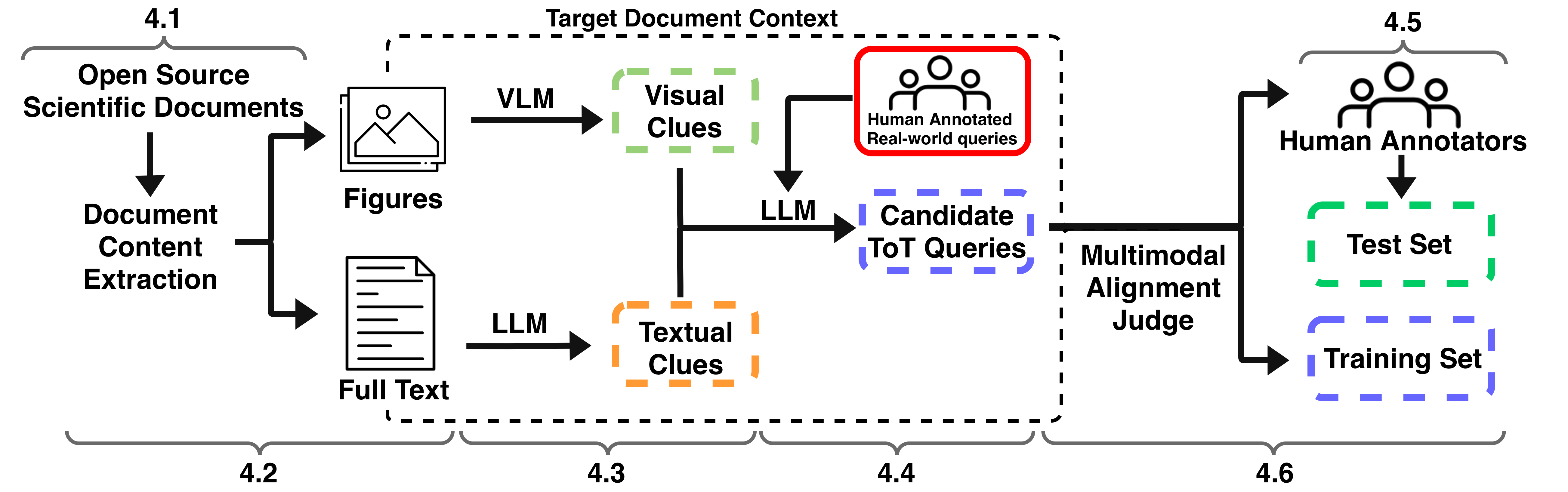}
    \caption{Pipeline for multimodal ToT query elicitation and validation for scientific documents. Numerical annotations refer to the sections discussing each element.}
    \label{fig:dataset_pipeline}
    \vspace{-0.5cm}
\end{figure}

In this section, we discuss the construction of the SciToT benchmark, following the pipeline shown in~\cref{fig:dataset_pipeline}. We extract visual and textual elements from scientific documents, generate clues and queries, and validate and filter the resulting queries through human annotation and an \ac{LLM} judge.

\subsection{Document Collection}
We construct our document collection from three sources and five scientific domains: Computational Linguistics from the ACL Anthology,\footnote{\url{https://aclanthology.org}} Medicine and Biology from PubMed Central (PMC),\footnote{\url{https://pmc.ncbi.nlm.nih.gov}} and Engineering and Physics from arXiv.\footnote{\url{https://arxiv.org}} We include only documents explicitly released under CC0 or CC BY licenses, permitting redistribution and reuse of documents, with CC BY requiring appropriate attribution~\cite{creativecommons_ccby, creativecommons_cczero}. The final collection consists of five sets of $20\text{K}+$ documents, resulting in a final corpus size of $100\text{K}+$ documents, and allows for evaluation across a range of different domains and visualizations. 

\subsection{Visual and Textual Element Extraction}
To generate clues and enable textual and visual retrieval methods, we extract visual and textual elements from PDF documents. Content extraction from documents is a substantial field of research~\cite{kim2022ocr, hu2024mplug, lv2023kosmos, wei2024vary, wei2024small,wang2024mineru}. Since content extraction is not the focus of this work, we rely on MinerU~\cite{wang2024mineru}, which satisfies the requirements of being open-source, has a low compute cost and is fine-tuned for scientific documents. For each document, we extract the text into markdown, and extract figures and tables into separate images. Qualitative inspection of extracted documents indicated that textual content generally preserved the source material and that figures and tables were generally correctly identified. 

\subsection{Memory Clue Extraction}
To facilitate query generation, we generate potentially remembered clues from the target document. Memory clues are extracted from both images and text. To ensure that clues are only related to a single modality and element, we separate clue extraction for text and images, following previous work on query generation~\cite{talmor2021multimodalqa}. 

For textual clues, we apply Qwen3-4B~\cite{yang2025qwen3}, providing it with instructions and the text of the paper, truncated to the token limit, following~\citet{he2025tip}. The model is instructed to output JSON-formatted text with a set of unstylized derived clues that are non-specific, and to exclude named entities or directly identifying information. For visual clues, we apply Qwen3-VL-4B-Instruct~\cite{bai2025qwen3}. The model is provided with a single image and is instructed to derive clues about the structural appearance as well as high-level semantic elements. 

\subsection{\ac{ToT} Query Elicitation}
\label{sec:tot_query_elicitation}
Our \ac{ToT} query elicitation generates a query from the set of clues of the target document. We apply Phi-4 14B~\cite{abdin2024phi}, using a different model provider from cue extraction to reduce single-provider bias and stylistic monotonicity~\cite{spiliopoulou2025play,bitton2025detecting,kim2025correlated,zhang2025invisible}. We generate queries based on both modalities, selecting the visual clues of at most 15 figures to stay within token limits. We closely follow the prompt style of~\citet{he2025tip}, providing the model with general role-play instructions, the clue set, a set of strict requirements (musts), and optional requirements (coulds). The prompt was adjusted to align with the scientific re-finding task, instructing the model to apply the clues exactly, and to follow the clue style elements described in~\cref{subsec:scitot_char}. We exclude instruction of the model to provide temporal or physical contextual clues, as these would require further complex user simulation, without adding a significant new challenge for retrieval. For visual elements, we instruct the model to focus on figures that stand out to it, as these are most likely to be remembered~\cite{konkle2010conceptual, Isola2011memorability}. All prompts are made available in our code repository.  

\subsection{Human Annotation}
\label{subsec:human_annotation}
To assess the quality of generated \ac{ToT} queries, we conduct a manual evaluation of 100 queries sampled from the test set with 20 queries per scientific domain. Two independent annotators with research backgrounds, neither involved in the development of the benchmark, evaluate all 100 query–document pairs. Annotators are presented with the generated query and the full source document and independently assess two dimensions: relevance, measuring the correspondence between the query and its source document, and well-formedness, assessing the clarity and interpretability of the query. Both dimensions are evaluated on three-point scales (0–2), corresponding to non-relevant, partially relevant, and highly relevant for relevance, and not well-formed, partially well-formed, and well-formed for well-formedness, aligning with previous work~\cite{kato2021test, roberts2021searching, wang2023scientific}. Annotators received written guidelines containing label definitions and illustrative examples and completed a calibration round using a separate set of examples covering all annotation categories during a one-hour training session. We describe annotator judgments and report the label distributions and inter-annotator agreement using weighted Cohen's $\kappa$ for each dimension. Annotator instructions are made available in our code repository.

\subsection{Filtering}
\label{subsec:filtering}
We systematically verify the alignment of the target documents with the generated query. We do so by applying LLM-as-a-Judge to verify whether the query components are supported by the document, and if the final query aligns with the document in general. We apply Gemma4-12B-IT~\cite{team2026gemma}, aligning with previous patterns of applying a different model family for evaluation~\citep{dietz2025principles}, as well as using a model with long-context capacity to ensure sufficient encoding capacity~\cite{tan2025judgebench}. The judge is supplied with the figures and full text of the target document, as well as the target query. The judge is instructed to assess each remembered component conservatively, and provide judgments with justifications; following this, it is instructed to take all components into account, and grade the target query on the same three-point scale as used by annotators. In filtering documents, we follow the approach of~\citet{weller2024followir}, discarding documents that do not align with the target. We discard all 0 relevance score elements for the training and test set, and report on the agreement with annotators. 

\begin{table}[t]
\centering
\footnotesize
\caption{Distribution of human annotation scores for relevance and well-formedness. Frequencies are averaged across 2 annotators. Linear-weighted Cohen's $\kappa$
measures inter-annotator agreement on paired items.}
\label{tab:annotation-results}

\begin{minipage}[t]{0.48\textwidth}
\centering
\textbf{Relevance ($\kappa = 0.210$)}\\[4pt]
\begin{tabularx}{\linewidth}{cp{3cm}r}
\toprule
\textbf{Score} && \textbf{Freq. (\%)} \\
\midrule
0 && 0.0 \\
1 && 7.5 \\
2 && 92.5 \\
\bottomrule
\end{tabularx}
\end{minipage}
\hfill
\begin{minipage}[t]{0.48\textwidth}
\centering
\textbf{Well-formedness ($\kappa = -0.034$)}\\[4pt]
\begin{tabularx}{\linewidth}{cp{3cm}r}
\toprule
\textbf{Score} && \textbf{Freq. (\%)} \\
\midrule
0 && 0.0 \\
1 && 5.5 \\
2 && 94.5 \\
\bottomrule
\end{tabularx}
\end{minipage}

\vspace{-0.5cm}
\end{table}

\begin{table}[t]
\centering
\footnotesize
\caption{Distribution of LLM-assigned relevance scores for
the annotated test subset and all saved LLM judgements. Linear-weighted Cohen's $\bar\kappa$ is
averaged across the two LLM--annotator comparisons (0.421 and 0.302).}
\label{tab:llm-annotation-results}

\begin{minipage}[t]{0.48\textwidth}
\centering
\textbf{Annotated Set ($\bar\kappa = 0.362$) \strut}\\[2pt]
\begin{tabularx}{\linewidth}{cp{3cm}r}
\toprule
\textbf{Score} && \textbf{Freq. (\%)} \\
\midrule
0 && 0.0 \\
1 && 3.0 \\
2 && 97.0 \\
\bottomrule
\end{tabularx}
\end{minipage}
\hfill
\begin{minipage}[t]{0.48\textwidth}
\centering
\textbf{Full Dataset \strut}\\[2pt]
\begin{tabularx}{\linewidth}{cp{1cm}rp{0.9cm}r}
\toprule
\textbf{Score} && \textbf{Freq.} && \textbf{Freq. (\%)} \\
\midrule
0 && 2 && 0.02 \\
1 && 369 && 3.21 \\
2 && 11,130 && 96.77 \\
\bottomrule
\end{tabularx}
\end{minipage}
\vspace{-0.5cm}
\end{table}

\section{Dataset Validation}
Validation of generated queries is carried out following the two components described in~\cref{sec:scitot_inf_need}. We first describe accuracy statistics for the Query dataset derived from human relevance annotations and LLM-as-a-Judge filtering, and then detail the statistical and qualitative alignment with tip-of-the-tongue queries. 

\subsection{Query-Document Alignment}
\cref{tab:annotation-results} summarizes human evaluations of 100 test queries. For relevance, 92.5\% of annotations were annotated as highly relevant, although inter-annotator agreement was limited, potentially reflecting the known sensitivity of Cohen's $\kappa$ to highly skewed label distributions \cite{feinstein1990high,byrt1993bias}. For well-formedness, 94.5\% of annotations received the highest score, but no meaningful chance-corrected agreement was observed. We attribute this to differences in annotators' preferences regarding query formulation, and limited difference in query formulation quality. 

In \cref{tab:llm-annotation-results} we show the LLM judge results, and agreement with human annotators. Overall, we observe the judge is slightly more lenient than human annotators, and assigns a score of 2 more often. We do find that it is better aligned with human judges than inter-human agreement, with higher agreement for both annotators. Overall, the automated judgments should primarily be interpreted as a filtering mechanism rather than a fine-grained measure of query quality. Across the full dataset, the judge identifies two non-relevant queries, which were manually confirmed and removed. The relatively modest agreement with human annotations limits the conclusions that can be drawn about the reliability of this filtering process. These findings support the overall quality of the generated queries while highlighting differences in evaluation judgments.

\subsection{ToT Query Alignment}
\begin{figure}[t]
    \centering
    \includegraphics[width=\linewidth, trim={0mm 5mm 0mm 0mm},
        clip]{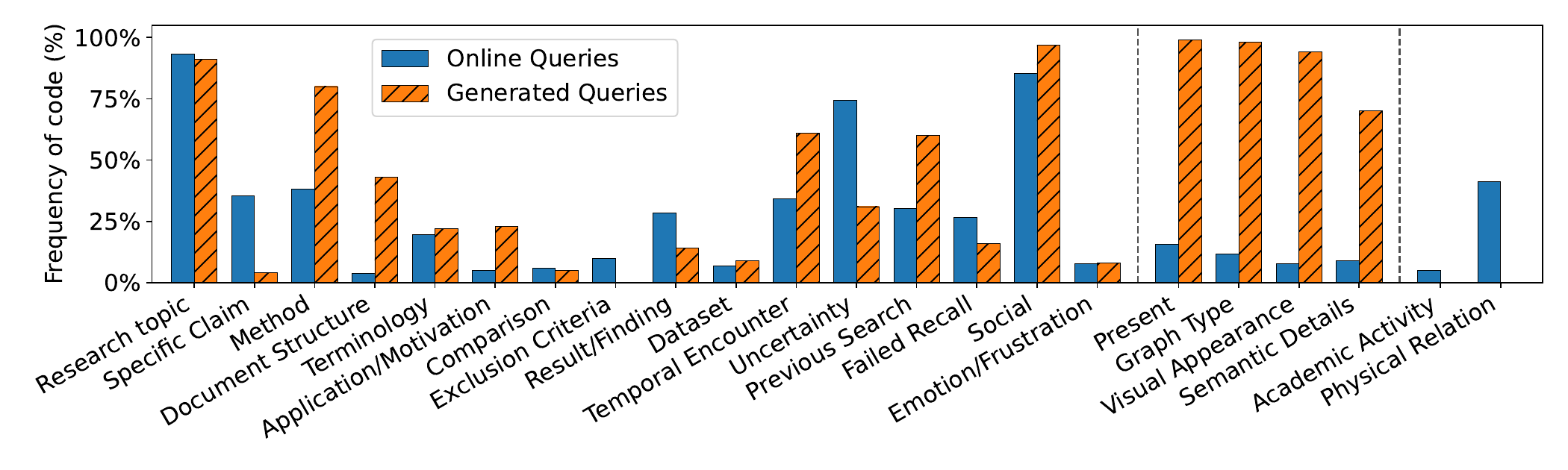}
    \caption{Distribution of code annotations found in online queries compared to those found in generated queries. Results are grouped into naturally generated, explicitly oversampled, and explicitly excluded items.}
    \vspace{-0.5cm}
    \label{fig:code_frequencies}
\end{figure}

We assess alignment between generated and real-world ToT queries using the annotation procedure described in~\cref{subsec:scitot_char}, applying the same codebook and annotators to $N=50$ generated queries. Figure~\ref{fig:code_frequencies} compares code frequencies across three groups separated by dashed lines. The rightmost group contains contextual elements deliberately excluded from generation, as they require modelling user history beyond paper content, as noted in~\cref{sec:tot_query_elicitation}. The middle group comprises visual elements, which are overrepresented by design to emphasize multimodal retrieval. The leftmost group contains elements moderated in frequency by the generation prompt, where we find high agreement, with an average $17.8\%$ difference in frequency. A notable difference includes that we observe fewer specific claims, likely because the selected sample of real-world queries included many mathematics-related requests for proofs. We also find more references to methods, document structure, and applications, which can be explained by the tendency of LLM-extracted clues to emphasize these aspects. Overall, generated queries reproduce much of the diversity and relative frequency of characteristics observed in real-world ToT queries.

\section{Retrieval Benchmark}
Scientific \ac{ToT} retrieval inherits several challenges from traditional \ac{ToT} retrieval, including incomplete recollections and uncertain descriptions~\cite{arguello2021tip,he2025tip}. These challenges are compounded in scientific literature search, where relevant information may be confined to a small portion of a lengthy document. To facilitate future work and provide insights into the challenges associated with our dataset, we provide implementations and evaluations of several baseline methods. 

We evaluate scientific \ac{ToT} retrieval as a known-item search task, where the objective is to retrieve the single target document from a collection based on an incomplete recollection query. Each scientific domain is evaluated independently, allowing future work to investigate both domain-specific retrieval effectiveness and cross-domain generalization. Since each query has a single target document, we report MRR@10, Recall@1, Recall@10, and Recall@100. 

\subsection{Retrieval Baselines}
We evaluate pretrained retrieval baselines across four
configurations: textual, visual, page-level, and multimodal
retrieval. For textual retrieval, we consider BM25~\cite{Robertson1994BM25}
for lexical matching, E5-small-v2~\cite{wang2022te5dense} and Qwen3-Emb-0.6B~\cite{yang2025qwen3} for dense semantic retrieval, and ColBERTv2~\cite{santhanam2022colbertv2}
for token-level late interaction. For visual and page-level
retrieval, we evaluate OpenCLIP~\cite{cherti2023reproducible}
and Qwen3-VL~\cite{bai2025qwen3}. For page-level retrieval, we also apply ColQwen2.5, a variant of the ColPali late interaction method for VLM retrieval~\cite{faysse2025colpali}. We additionally evaluate Qwen3-VL for multimodal retrieval using both extracted text and figures. All exact model versions and implementation details can be found in our public repository.

All methods are evaluated without fine-tuning.
Textual methods use extracted paper text, truncated to satisfy model input limits. OpenCLIP encodes individual figures or pages, using maximum cosine similarity for relevance scores. For Qwen3-VL we jointly encode all available inputs for each configuration, truncated to the token limit, producing a single document-level embedding to allow for in-document context. ColQwen2.5 is designed as a page-level method, to retain this architecture, we apply individual page embeddings and use the max-similarity in the ranking stage~\cite{faysse2025colpali}. To reduce the size of the index to a feasible size for our storage budgets, we apply built-in hierarchical token pooling.
\subsection{Benchmark Results}

\begin{table}[t]
\centering
\footnotesize
\caption{Retrieval effectiveness across methods and modalities.
Each cell is an unweighted mean over available results from the
five retrieval datasets; missing values are excluded.
Index size (decimal GB) is averaged over datasets.}
\setlength{\tabcolsep}{1pt}
\begin{tabularx}{\textwidth}{@{}l l *{4}{>{\centering\arraybackslash}X} r@{}}
\toprule
\textbf{Method} & \textbf{Modality} &
\textbf{MRR@10} & \textbf{R@1} &
\textbf{R@10} & \textbf{R@100} &
\textbf{Index Size} \\
\midrule
BM25 & Text & 0.123 & 0.095 & 0.194 & 0.371 & 0.113 \\
E5 & Text & 0.128 & 0.090 & 0.228 & 0.451 & 0.033 \\
ColBERTv2 & Text & 0.115 & 0.098 & 0.150 & 0.197 & 0.144 \\
Qwen3-Emb & Text & 0.710 & 0.653 & 0.825 & 0.914 & 0.089 \\
\midrule
OpenCLIP & Visual & 0.035 & 0.020 & 0.078 & 0.188 & 1.093 \\
Qwen3-VL & Visual & 0.637 & 0.569 & 0.775 & 0.903 & 0.177 \\
\midrule
OpenCLIP & Page & 0.010 & 0.005 & 0.024 & 0.091 & 0.557 \\
Qwen3-VL & Page & 0.793 & 0.746 & 0.881 & 0.932 & 0.178 \\
ColQwen2.5 & Page & 0.638 & 0.584 & 0.763 & 0.858 & 78.982 \\
\midrule
Qwen3-VL & Text\,+\,Visual & 0.750 & 0.699 & 0.843 & 0.899 & 0.178 \\
\bottomrule
\end{tabularx}
\label{tab:retrieval_results}
\vspace{-0.6cm}
\end{table}


In~\cref{tab:retrieval_results} we show the retrieval effectiveness results. For textual methods, we observe that Qwen3-Emb substantially outperforms traditional retrieval baselines. Textual baselines BM25, E5, and ColBERTv2 achieve relatively low effectiveness (MRR@10 of 0.115--0.128), compared to 0.710 for Qwen3-Emb using text alone. These results may be explained by noise in queries, limited context length (512 tokens for E5, and 32 query tokens for ColBERTv2). For visual methods, we find poor performance for OpenCLIP, indicating that visual information alone is insufficient without representations capable of capturing more complex semantic relationships. Notably, we find that Qwen3-VL performs relatively well using only image elements, indicating that visual query descriptions can be a useful retrieval signal in this task. Page-level retrieval with Qwen3-VL achieves the highest effectiveness (MRR@10 of 0.793). Furthermore, including both text and visual modalities also improves retrieval effectiveness for Qwen3-VL. For ColQwen2.5, we find that it performs worse than Qwen3-VL, and requires significantly more index storage, which aligns with retrieval results reported in previous work~\cite{khalighinejad2026document}. We do note that Qwen3-based models are used for both clue extraction and retrieval; it is possible that representational biases may favor these retrievers, although the query generation model is different. 

Our findings highlight that traditional retrieval methods can struggle with scientific \ac{ToT} retrieval, and reveal opportunities for developing efficient yet fine-grained multimodal retrieval methods that align uncertain memory cues with localized document content. Furthermore, data augmentation methods such as query rewriting, document summarization, or image description extraction may lead to improvements for methods which are limited by context length. 

\section{Resource Availability and Intended Use}
\label{sec:availability}

We release SciToT on Hugging Face with a persistent DOI, including source documents, extracted textual and visual content, memory cues, queries, judgments, and evaluation splits.\footnote{\url{https://doi.org/10.57967/hf/10991}} The accompanying datasheet~\cite{10.1145/3458723} documents the resource's composition, collection and preparation procedures, intended uses, and maintenance policy. Furthermore, we release the dataset generation pipeline\footnote{\url{https://github.com/stanfris/multimodal-tip-of-tongue-retrieval-for-scientific-papers}} and retrieval implementations,\footnote{\url{https://github.com/stanfris/multimodal-tot-retrieval}} which support further development and dataset downloading. All generated resources and code are released under MIT licenses, while source documents retain their original CC0 or CC BY licenses.

SciToT supports the development and evaluation of scientific \ac{ToT} retrieval methods, particularly multimodal retrieval from noisy and incomplete recollections. The generation pipeline also enables the construction of related benchmarks in other tasks or domains. The resource is intended for permanent availability, with updates managed through the public repositories.

\section{Conclusion} 
We have introduced SciToT, a benchmark for scientific \ac{ToT} retrieval, comprising over 100K scientific documents across five domains. We have proposed a scalable query elicitation pipeline that combines textual and visual memory cues to generate queries reflecting incomplete recollections of scientific papers. Human evaluation has demonstrated high query–document alignment and well-formedness, while an empirical comparison with real-world scientific re-finding requests showed that the generated queries capture many of their characteristics.

Our retrieval experiments highlighted the challenges of scientific \ac{ToT} retrieval. Conventional lexical and dense retrieval baselines achieved limited effectiveness, but more expressive document representations improved retrieval performance, highlighting the importance of matching remembered information to localized content within scientific documents and capturing visual elements.

The synthetic nature of the queries in the SciToT benchmark and their deliberate emphasis on visual clues may limit their representativeness of naturally occurring requests. Users of SciToT should be mindful of these limitations when training or fine-tuning their retrieval models. We encourage future work on expanding the coverage and resources of SciToT, in terms of both \ac{ToT} query (sub)types and additional scientific domains. We also call for research on generalizability across these two dimensions.


%
%
\bibliographystyle{splncs04nat}
\bibliography{references}
\end{document}